\documentclass[runningheads]{llncs}
\usepackage[T1]{fontenc}
\usepackage{graphicx}

\usepackage{hyperref}
\usepackage{amsmath}
\usepackage{amssymb}
\usepackage{xcolor}
\usepackage[english]{babel}
\usepackage{tcolorbox}

\usepackage{adjustbox}
\usepackage{booktabs}
\usepackage{multirow}
\usepackage{colortbl}

\usepackage[show]{chato-notes}
\usepackage{enumitem}

\setitemize{noitemsep}
\setlist{topsep=0pt, leftmargin=*}
\newcommand{\uls}{\begin{itemize}[leftmargin=*]}
\newcommand{\ule}{\end{itemize}}
\newcommand{\ols}{\begin{enumerate}[leftmargin=*]}
\newcommand{\ole}{\end{enumerate}}
\newcommand{\li}{\item}

\newcommand{\llm}{\phi_{\text{LLM}}}
\newcommand{\topk}{\theta(q)_k}
\newcommand{\bs}{F_{\text{BERT}}}

\newcommand{\cm}[1]{\textcolor{black}{#1}}

\newcommand{\fzA}[1]{\textcolor{black}{#1}}

\newcommand{\para}[1]{\paragraph{\textnormal{\textbf{#1.}}}}

\usepackage{marginnote}
\newcommand{\pageenlarge}[1]{\marginnote{#1}\enlargethispage{#1\baselineskip}}

\renewcommand{\vec}[1]{\mathbf{#1}}

\begin{document}
\title{Is Relevance Propagated from Retriever to Generator in RAG?}
\titlerunning{Is Relevance Propagated from Retriever to Generator in RAG?}
%
\author{Fangzheng Tian\orcidID{0009-0000-3282-0220} \and
Debasis Ganguly\orcidID{0000-0003-0050-7138} \and
Craig Macdonald\orcidID{0000-0003-3143-279X}}
\authorrunning{Tian et al.}
%
\institute{
University of Glasgow, Glasgow, UK
\\
\email{f.tian.1@research.gla.ac.uk},
\email{debasis.ganguly@glasgow.ac.uk},
\email{craig.macdonald@glasgow.ac.uk}
}
\maketitle              
\begin{abstract}
Retrieval Augmented Generation (RAG) is a framework for incorporating external knowledge, usually in the form of a set of documents retrieved from a collection, as a part of a prompt to a large language model (LLM) to potentially improve the performance of a downstream task, such as question answering.
Different from a standard retrieval task's objective of maximising the relevance of a set of top-ranked documents, a RAG system's objective is rather to maximise their total \textit{utility}, where the utility of a document indicates
whether including it as a part of the additional contextual information in an LLM prompt improves a downstream task.
Existing studies investigate the role of the relevance of a RAG context for knowledge-intensive language tasks (KILT), where relevance essentially takes the form of answer containment. In contrast, in our work, relevance corresponds to that of topical overlap between a query and a document for an information seeking task.
Specifically, we make use of an IR test collection to empirically investigate whether a RAG context comprised of topically relevant documents leads to improved downstream performance.
Our experiments lead to the following findings: (a) there is a small positive correlation between relevance and utility; (b) this correlation decreases with increasing context sizes (higher values of $k$ in $k$-shot); and (c) a more effective retrieval model generally leads to better downstream RAG performance.
\end{abstract}
\section{Introduction}

Retrieval Augmented Generation (RAG) is a framework to help Large Language Models (LLMs) generate more accurate answers for tasks that require domain-specific knowledge~\cite{ragInKnowledgeIntensiveNLP}. This knowledge is passed to an LLM \fzA{as context} - \cm{usually in the form of a} ranked list of retrieved documents, which are then inserted in a prompt template designed for a specific downstream task \cite{dong2024surveyincontextlearning,ragsurvey}. Usually, these retrieved documents are topically similar to the input instance, e.g., for a particular input question in a question-answering (QA) task, they can be constituted of semantically similar questions along with their answers as retrieved from a collection \cite{demoRetrieveInICL}.

A challenging problem in RAG is to find an effective way to construct this context for each input (query)~\cite{powerOfNoise,lostInTheMiddle}. In contrast to the standard ranking task, where the aim is to maximise the likelihood of relevance of the retrieved documents, the task in RAG is to maximise the likelihood of the usefulness of this retrieved context for a downstream task \cite{iclPerspective}.

While existing research has investigated the role of relevance in generating quality output for QA systems \cite{powerOfNoise,lostInTheMiddle,zamani-rag}, the notion of relevance in their work is largely different from the notion of topical relevance required in an information seeking task (e.g., query-focused summarisation \cite{evalGenIR}), where relevance is obtained with manual assessments via depth-pooling \cite{Cormack98,TREC-DL-2019-overview,DBLP:conf/sigir/GangulyY23,Sanderson10}.
More precisely speaking, the notion of `relevance' in existing studies applies specifically to knowledge-intensive language tasks (KILT benchmark \cite{petroni-etal-2021-kilt}), e.g., the CommonQA \cite{talmor-etal-2019-commonsenseqa} or the NaturalQA benchmarks \cite{kwiatkowski-etal-2019-natural}, where
a Boolean indicator determines whether the correct answer is present in a retrieved context (a passage of text from which the correct answer needs to be extracted out - commonly called `provenance' in these benchmarks).

Different from the existing studies for \textbf{KILT tasks}, \fzA{where the \cm{relevance} is evaluated} in the form of \textbf{answer containments}
\cite{powerOfNoise,lostInTheMiddle,zamani-rag}, in our paper, \fzA{\textbf{relevance of retrieved context}} corresponds to the \textbf{topical match} between a query and \fzA{the included} documents for an \textbf{information seeking task}, i.e., the standard IR sense of the term. \fzA{Inspired by }the work of \cite{benchmarkOfAnswerEvaluationFromRetriever}, we propose an experiment setup for RAG using an IR test collection, where a set of queries is used as input questions to an LLM - the task is to generate an answer that satisfies the information need underlying the question \cite{evalGenIR}. The quality of the generated answer is then measured by its similarity with the set of assessed relevant documents for the query (which is a part of the IR test collection).

A major difference of our work from \cite{benchmarkOfAnswerEvaluationFromRetriever} is that while they investigated only the 0-shot answer generation (no context is involved), we investigate the role of retrieved contexts for effective answer generation in a $k$-shot setup.
We provide a precise definition of the usefulness of a retrieved context as \fzA{the relative change in answer effectiveness induced by including the corresponding context.} \cm{This} \textbf{utility measure} \fzA{follows the general counterfactual attribution methodology, i.e. quantifying the varying inputs’ influence upon a model’s outputs (as used in e.g. model explanations~\cite{DBLP:conf/sigir/VermaG19,verma2022counterfactualexplanationsalgorithmicrecourses}).}
More specifically, we investigate the relation between the relevance of the ranked list of $k$ documents in $k$-shot RAG and the quality of the generated answer as measured by its semantic similarity with the known relevant documents for a query~\cite{benchmarkOfAnswerEvaluationFromRetriever}.
In other words, we are interested in finding out if a context comprised of more relevant documents can lead to generating more relevant answers, or is the \textbf{relevance lost in transmission} across the generation phase as the data flows from the `retriever' to the `generator' components of a RAG system?

\begin{figure}[t]
\centering
\includegraphics[width=0.9\textwidth]{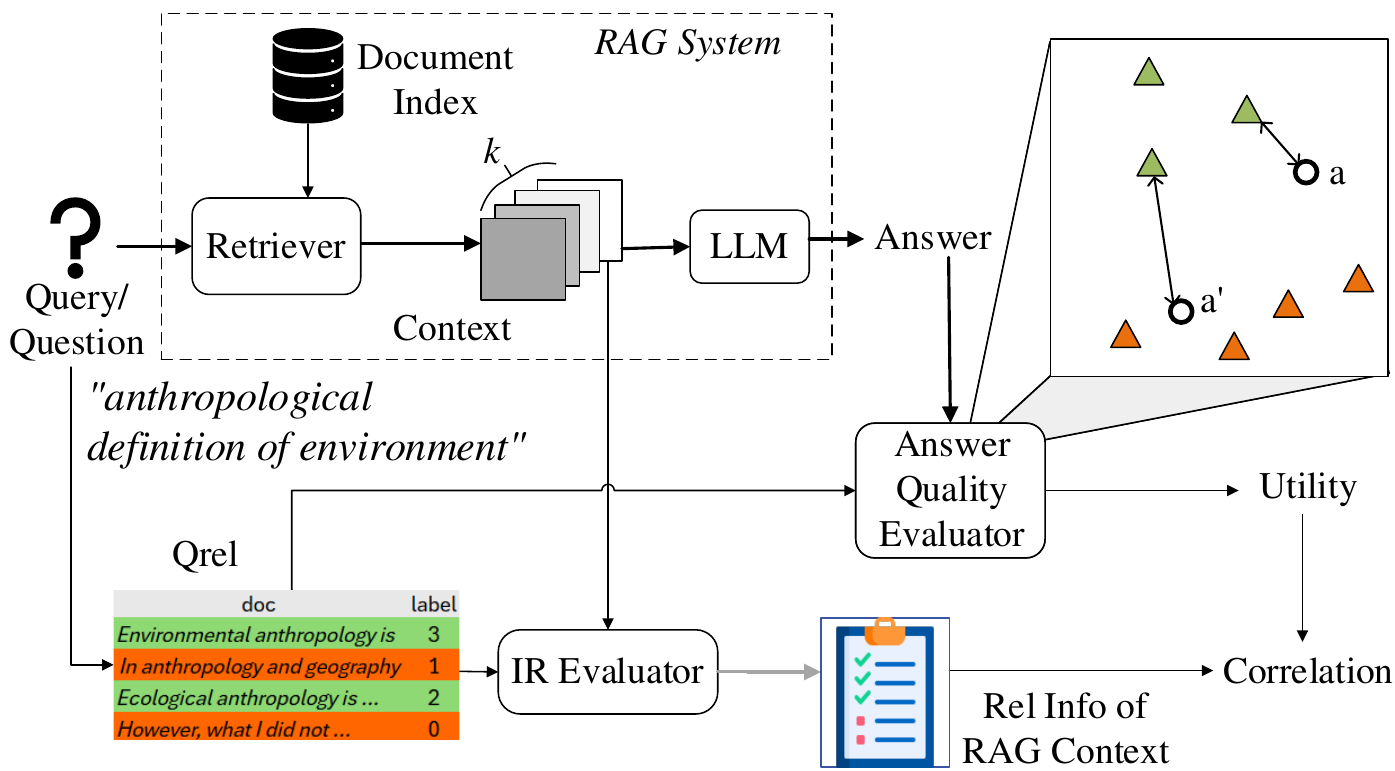}
\caption{The illustration of the generator (LLM) with $k$ top-retrieved documents comprising a RAG context. Conducting experiments with an IR test collection enables the `retriever' component to be evaluated with standard IR metrics using available ground truth of known relevance assessments. The `generator' component is evaluated by computing the similarity of a generated answer's embedding w.r.t. the embeddings of the judged relevant documents (shown as the green triangles). For instance, \fzA{generated answer $\vec{a}$ has a shorter distance to its closest relevant document than
$\vec{a}'$ does, so $\vec{a}$ gets a higher $\bs$ than $\vec{a}'$}. 
\label{fig:utility_in_QA_task}
}
\end{figure}

Figure \ref{fig:utility_in_QA_task} presents a schematic overview of our proposed evaluation framework to investigate the effect of RAG contexts' topical relevance on the quality (measured in terms of relevance) of the generated answers.
More details about the evaluation framework are presented in Section \ref{sec:measure}.

\begin{contribution}
The overall contribution of our paper is the proposal of a novel experiment framework for RAG systems, in which:
\uls 
\li We introduce a measure for context utility that focuses on the relative gain brought by the retrieved context wrt. the 0-shot answer quality;
\li \cm{We correlate the relevance of the retrieved context (e.g.~nDCG) and their} utility in downstream RAG tasks and use this correlation to indicate whether relevant information is effectively transmitted from the retrieval results to the generated answers.
\ule
\looseness -1 \cm{In our experiments examining the relevance-utility correlation, we obtain interesting findings from the experiments based on our framework, including:}
\uls
\li The correlation between nDCG and the corresponding utility is positive but small, which means that \textit{a part of relevance is indeed lost in transmission} from the retrieval result to the generated output;
\li This correlation decreases as the context size increases and is relatively insensitive to the order of the documents in the context for the tested LLM.
\ule
\end{contribution}

\fzA{To help further studies in this domain and the reproduction of the results, we make our experiment framework implementation publicly available at: \\\url{https://github.com/DanielTian97/rag-utility.git}}

\section{Related Work}

\para{RAG background} 
The concept of retrieval-augmented generation (RAG) was first introduced in \cite{ragInKnowledgeIntensiveNLP}. Subsequent studies have adapted it for a number of different applications, e.g., for improving the performance of knowledge-intensive language tasks (KILT) \cite{petroni-etal-2021-kilt} which include question answering \cite{kwiatkowski-etal-2019-natural,talmor-etal-2019-commonsenseqa}, hallucination and factual incorrectness mitigation \cite{FLARE}, summarisation \cite{edge2024localglobalgraphrag}, software code generation \cite{li2023acecoderutilizingexistingcode}. For a more comprehensive review on RAG literature, see \cite{ragsurvey,ragReview}.

Optimal context selection plays an important role in determining RAG's performance. While the most common approach to select this context is based on lexical or semantic similarities with the input text, e.g., question, instruction, query etc. \cite{liu-etal-2022-makes,demoRetrieveInICL}, other selection methodologies with different objectives have also been explored, e.g., \cm{for example selecting diverse documents as retrieved context}~\cite{levy-etal-2023-diverse},
or learning a task-specific ranking model to maximise the likelihood of the \fzA{retrieved} 
context improving the downstream performance \cite{rubin-etal-2022-learning}.
It has also been empirically verified that RAG performance depends not only on the set of documents selected as the context, but also on the order in which these are appended to the input text of an LLM \cite{DBLP:conf/acl/KumarT21,lu-etal-2022-fantastically}.

Existing research, however, has not explicitly investigated the \textit{effect of topical relevance of the RAG context} on the \textit{quality of the generated output} (in our case, an answer), which is the core objective of this paper.

\para{Analysing the influence of documents on downstream RAG performance}

Sauchuk et al. \cite{roleOfRelevanceInNLP} \cm{reported} that topically non-relevant documents lead to substantial degradation of downstream performance of cascaded IR-NLP systems (precursor to an end-to-end RAG system but characteristically similar to it). {More recently,} \cite{powerOfNoise,lostInTheMiddle} also showed that not only does the presence of relevant documents lead to improved performance in the downstream tasks compared with 0-shot generation, but the order of those relevant documents also affects the performance.
In particular, the authors of \cite{lostInTheMiddle} reported that
``\textit{model performance is highest when relevant information occurs at the beginning or end of its input context}'' and coined the term "lost in the middle". Their observation was further corroborated by the work of \cite{powerOfNoise}, with some further analysis on the effects of non-relevant vs.\ randomly sampled off-topic documents.

\looseness -1 In our work, we explore if the observations reported in \cite{powerOfNoise,lostInTheMiddle} for a notion of relevance restricted to \textit{answer containment} are consistent across topical relevance in the form of manually assessed ground truths. Similar to \cite{zamani-rag}, we also report the correlation between retrieval effectiveness and downstream RAG performance.

\para{Evaluation Methods of RAG Performance}

A RAG model's performance measure depends on the specific downstream tasks \cite{surveyLLMEvaluation}, e.g., in QA or summarization tasks, the common evaluation metrics measure the overlap between the ground-truth and a generated output either at a lexical level or at a semantic level. Lexical measures include
METEOR~\cite{metero}, ROUGE~\cite{rouge}, BLEU~\cite{bleuScore}, whereas semantic ones include BERTScore~\cite{finegrainedAnalysisOfBERTScore,bertScore}. Other overlap measures have also been proposed, e.g., FActScore \cite{factScore_EMNLP23} employs aspect-based matching to measure the factual correctness of generated answers, whereas the studies reported in \cite{huang2024empiricalstudyllmasajudgellm,zhuo2024icescoreinstructinglargelanguage} used LLMs to evaluate the quality of generated output.

In contrast to evaluating factual correctness, the authors of \cite{benchmarkOfAnswerEvaluationFromRetriever} proposed a \textit{relevance-based} evaluation strategy for measuring the quality of generated answers by conducting their experiments on an IR dataset where relevance ground-truths are available. Different from evaluation methodologies that compute the overlap of a generated output with a ground-truth one, the authors of \cite{benchmarkOfAnswerEvaluationFromRetriever} used the centroid of the relevant documents for a query to construct a pseudo-ground truth output - an approach that we also adapt in our work.

\looseness -1 \para{Research Gap} Overall, the closest prior work concerned with the relation between RAG context and RAG performance are \cite{powerOfNoise,lostInTheMiddle}, but neither directly investigated the problem with a conventional IR definition of topical relevance, as is addressed in our work. Moreover, those studies focused on overall RAG performance; in contrast, our work distinguishes the LLM's inherent performance in the downstream RAG task and the relative change brought by the retrieval context, which we define as utility. To investigate this problem systematically, we extended the framework that leverages IR benchmarks to evaluate RAG~\cite{benchmarkOfAnswerEvaluationFromRetriever}. We introduce our RAG Utility evaluation framework next.

\section{Evaluating Relevance vs. Utility in RAG}\label{sec:measure}

In this section, we formally define a RAG evaluation framework that uses an IR test collection to measure the relation between the relevance of retrieved documents and that of the output generated in a downstream task.

\subsection{Performance Measure of a Downstream Generation Task} \label{ss:dp}
As the first step in our evaluation framework, we define the performance measure of a RAG downstream task.
Similar to the experimental settings of \cite{benchmarkOfAnswerEvaluationFromRetriever}, we use the passage-level relevance information of queries in an IR test collection. Formally, we can define an IR test collection as $\langle Q, \mathcal{C}, R_Q \rangle$, where $Q$ is a set of queries, $\mathcal{C}$ is a collection of documents, and $R_Q$ is the set of available relevance assessments (qrels) of the queries.

The RAG task in this setup corresponds to \textit{generating} an answer $a_q = A(q; \llm, \topk)$ that seeks to provide relevant information to the query $q \in Q$, where the answer generation function $A$ depends on: (a) an LLM model $\llm$ with frozen parameters, and (b) contextual information, $\topk$, comprised of a list of top-$k$ documents retrieved by an IR model $\theta$. 
In contrast to a standard IR task, where the objective is to retrieve the relevant documents $R_q$ for query $q$, the objective here is to generate a single answer that aligns with the relevant information contained in the set of documents $R_q$.  

Similar to \cite{benchmarkOfAnswerEvaluationFromRetriever}, we define the performance measure of an answer $a_q$ as
\begin{equation}
\mathcal{P}(a_q) = \max_{r \in R_q} \sigma(\vec{a}_q, \vec{r}) \label{eq:downstream-performance},    
\end{equation}
where $\sigma$ is an abstract similarity function, and $\vec{a}_q$ and $\vec{r}$, respectively, denote the embedded representations of the answer text $a_q$ and a relevant document text $r \in R_q$, with $R_q$ being the set of assessed relevant documents for $q$.

In particular, we employ BERTScore \cite{bertScore} as the similarity function $\sigma$, as prescribed in \cite{benchmarkOfAnswerEvaluationFromRetriever}. The BERTScore measure between a pair of text is a semantic extension of the BLEU score \cite{bleuScore} metric that, instead of aggregating their term overlaps, rather aggregates the cosine similarities between their constituent token embeddings. Following the standard practice, we work with the F-score version from the BERTScore paper \cite{bertScore}, which we refer to as $\bs$ from here on.

Although previous work \cite{walert} has shown that BERTScore may not be entirely effective for out-of-domain question answering, our experiments on an IR test collection only involve in-domain queries, with each query having at least one relevant document. 

\subsection{Utility of Contextual Information}

We now define the utility of contextual information as the relative change in the performance measure (Equation \eqref{eq:downstream-performance}) between a non-contextual (i.e., 0-shot) generation and a contextual ($k$-shot RAG) one.
Formally,
\begin{equation}
U(\topk) = \frac{\mathcal{P}(A(q; \llm, \topk)) - \mathcal{P}(A(q; \llm, \varnothing))}{\mathcal{P}(A(q; \llm, \varnothing))},
\label{eq:utility}
\end{equation}
where $\theta: q \mapsto \{d_1,\ldots, d_k\}$ is a retrieval model that, given a query $q$, retrieves a ranked list of top-$k$ documents, which we denote by $\topk$.

Intuitively speaking, the notion of utility as defined in Equation \eqref{eq:utility} captures the usefulness of a set of top-retrieved documents relative to the 0-shot performance (an LLM's inherent capability of generating relevant output). A positive value of utility indicates that the contextual information retrieved from a collection improves the answer quality over 0-shot generation, and, conversely,  a negative utility value indicates that the retrieved context information is detrimental to 0-shot answer generation quality. A reason to work with relative gains instead of absolute ones is that it can correctly capture the improvements over ineffective 0-shot predictions.

\subsection{Measuring Relevance Transmission from Retriever to Generator} \label{ss:eval_metrics}

To see how much of relevant information actually gets transmitted via the generation process, we propose to measure the correlation between the utility values (Equation \eqref{eq:utility}) and the relevance of the retrieved context comprised of the top-$k$ documents retrieved for a query $q$. Formally speaking, we measure
\begin{equation}
\rho(U(\theta(Q)_k), \mu(\theta(Q)_k)),\,\,
\text{where}\,\, \theta(Q)_k = \cup_{q \in Q} \topk,
\label{eq:correlation}    
\end{equation}
with the notations being explained as follows.
\uls
\li $U(\theta(Q)_k)$ denotes a set of utility values computed for each top-retrieved set of documents $\topk$;
\li $\mu(\theta(Q)_k))$ denotes a set of IR performance measure values (e.g., nDCG, which we use for our experiments) computed for the contextual information $\topk$ retrieved for each query $q$ in $Q$;
\li $\rho$ denotes a correlation metric (e.g. Pearson's $r$ in our experiments).
\ule

The correlation between the relevance measured at the retriever and at the generator, as shown in Equation~\eqref{eq:correlation}, indicates how effectively an LLM can translate the relevant context retrieved from a collection of documents into a generated answer text that is also relevant to the question.

This correlation measure of Equation~\eqref{eq:correlation} depends on several explicit and latent factors. While the explicit factors include the LLM being used, the retrieval model $\theta$, and the RAG context size $k$, the more latent factors may include the specificity of the information needs of the questions etc. Our research questions, as described in the next section, investigate how the more explicit factors, namely the retrieval model $\theta$ and the RAG context size $k$, affect the correlation between the relevance in the retrieved context and the relevance in the generated answer.

\section{Experiment Setup} \label{sec:experiment}

\para{Research Questions}

Our first research question is about analysing the effects of the retrieval model $\theta$ on the downstream utility in Equation \eqref{eq:utility}. The second research question is about the correlation as described in Equation \eqref{eq:correlation}. Usually, retrieval models have different levels of abilities in ranking documents based on their estimated relevance to a topic. In particular, we study one lexical (unsupervised) and another neural (supervised) model - specifically, BM25 and MonoT5~\cite{monot5}.
\cm{These models are compared with two extreme cases: an oracle and and an {\em adversarial} oracle. Specifically,} by using the available relevance information $R_q$ for each query $q$, the oracle constructs a completely relevant RAG context by sampling $k$ \textit{judged relevant} documents from $R_q$, while the adversarial oracle constructs a completely non-relevant RAG context by sampling $k$ \textit{judged non-relevant} documents. This setup leads to the following research questions.

\uls
\li \textbf{RQ-1}: Is the contextual information of a RAG system obtained with a more effective ranking model
of higher utility (i.e., better improvements over 0-shot generation) than that of less effective ones?

\li \textbf{RQ-2}: How does the correlation between the relevance of the RAG context and the downstream utility change with the number of documents constituting the context?

\ule

\para{Datasets}
To study the effects of relevance on downstream RAG performance using an IR-based experiment setup, we employ the TREC DL'19~\cite{TREC-DL-2019-overview} and DL'20~\cite{TREC-DL-2020-overview} test collections, constituting
$43$ and $54$ queries, respectively, with depth-pooled manually assessed relevance judgements. The underlying document collection for the TREC DL queries is the MS MARCO passage collection, which contains $8.8$ million passages~\cite{MSMARCO-DATASET}.

In addition to evaluating our results on TREC DL (deep relevance assessments over a small number of queries), 
we also conduct experiments on the MS MARCO dev\_small queryset, which has a much larger number of queries (6980 in total), but with much shallower relevance assessments.

\para{LLM and Prompt Details}

Our experiments were conducted on
an 8-bit quantised instruction-tuned model of
Llama-3-8b\footnote{\url{https://huggingface.co/QuantFactory/Meta-Llama-3-8B-GGUF}}~\cite{llama3}.
To account for the stochastic nature of the generation process of an LLM, we conduct our analysis on an average across 5 runs.

The prompt template used in our experiments instructs an LLM to generate a single answer to a question (TREC DL query) as per both its internal knowledge and the contextual information provided to aid the generation.
Figure \ref{fig:prompt} shows the prompt used in our experiments along with an example query and BM25-retrieved documents about it. The average length of each document in the MS MARCO passage collection is about 56 words, and we found that
$16$ such documents fit within 
a $2048$-token input buffer.
We thus set the maximum value of the RAG context size to 15 in our experiments.

\begin{figure}[t]
\centering
\begin{tcolorbox}[size=small]
\textbf{You are an expert at answering questions based on your own knowledge and 
related context. Please answer this question based on the given 
context. End your answer with STOP.}
\\ 
\textbf{Context 1:} \{Definition of SIGMET. plural. 
... before takeoff — compare airmet.\}\\
\textbf{Context 2:} \{sigmet SIGMET, or Significant 
... convective and non-convective.\}\\
\textbf{Question:} \{definition of a sigmet\}\\
\textbf{Now start your answer.} \\
\textbf{Answer:} 
\end{tcolorbox}
\caption{The prompt template (shown bold-faced) used in our experiments with example values substituted for the placeholders (shown enclosed within \{\}). The query shown is a sample from the TREC DL'19 queryset, and the RAG context is comprised of the top 2 documents retrieved with BM25.
\label{fig:prompt}
}
\end{figure}

\para{IR Models Investigated}

To obtain the RAG contexts, we employ a lexical ranker (namely BM25) and a supervised neural ranker (namely MonoT5 \cite{monot5}), \fzA{as provided by PyTerrier~\cite{pyterrier}}. The documents used as context constitute the top-$k$ results of a retrieval model $\theta$ at different cut-offs to yield different context sizes (denoted as $\topk$ in Equations~\eqref{eq:downstream-performance} and \eqref{eq:utility}).
More details about the IR models \cm{and oracles} follow.
\begin{itemize}
\item \textbf{BM25}: A lexical retriever based on a variant of TF-IDF-based term weighting coupled with document length normalisation. The $k$ top-ranked documents are appended to the RAG context (see Figure \ref{fig:prompt}) in decreasing order of retrieval score, i.e., the top-ranked document is `Context 1', and so on.
    
\item \textbf{MonoT5}: A cross-encoder re-ranker based on the contextual embedding of query-document pairs~\cite{monot5}. In our setup, the MonoT5 top-$k$ were obtained by re-ranking the top-100 BM25 results. Similar to BM25, the context documents are appended to the LLM prompt in descending order of retrieval score.

\item \textbf{BM25-r} and \textbf{MonoT5-r}: For both the aforementioned IR models, to investigate the potential adverse effects of not conforming to the order induced by the retrieval scores,
we construct the RAG context by appending each document from the top-$k$ list $\topk$ in ascending order of retrieval score. 
In this way,
the most likely relevant documents are put at the end of the context, i.e., the top-ranked document is mapped to `Context $k$' (see Figure \ref{fig:prompt}). We denote these contexts with an `r' as the suffix.

\item \textbf{Rel (Oracle)}: To study the upper bound on the RAG performance, we experiment with an oracle approach that samples a set of $k$ relevant documents from the known set of assessed documents, thus appending them to the LLM prompt in an arbitrary order. If the number of relevant documents for a query is less than $k$ then we take all of them.

\item \textbf{NRel (Adversarial Oracle)}: In contrast to the relevant-only approach, we also study the worst-case settings by including a sample of $k$ non-relevant documents (assessed non-relevant documents from the qrels file of TREC DL'19 and TREC DL'20) into the RAG context.

\end{itemize}

\noindent We do not explore oracles for the dev\_small queryset \cm{due to the shallowness of the relevance assessments (only 1.1 relevant documents per query, on average).} 

\looseness -1 \para{Evaluation Metrics}
As described in Section \ref{ss:eval_metrics}, we measure the quality of the retrieved list of $k$ documents constituting a RAG context with nDCG@$k$. The downstream performance is measured using \fzA{the F1-based BERTScore}~\fzA{\cite{bertScore}} between the generated answer and the known relevant documents (denoted $\bs$, see also Figure \ref{fig:average_utility}).
We employ two different settings for constructing the ground-truth of relevant documents for the TREC DL query sets. In the first setting, following the standard practice of IR evaluation, we consider the documents with relevance labels ($r_{\text{DL}}$) of 2 and 3 to compute $\bs$. In the second setting, we apply a more restrictive notion of relevance where we consider only the highly relevant documents ($r_{\text{DL}}=3$) for $\bs$ computation. Only one configuration is available for the MS MARCO dev\_small queryset as the relevance labels are binary (results shown as $r_{\text{dev\_small}}=1$ in Table \ref{table:statistics}).

\begin{table}[t]
\centering
\caption{Average utilities of different RAG contexts retrieved by BM25 and MonoT5 (denoted MT5), including their reversed-order (-r) variants, and `relevant-only' (Rel) and `non-relevant only' (NRel) for context sizes 2, 5, 10 and 15.
To compute the relative gains in the $\bs$ scores, we use two configurations for TREC DL: (1) the documents with relevance labels ($r_{\text{DL}}$) of 2 and 3 are included in the computation; and (2) only the documents with label 3 are considered relevant.
A $\dagger$ alongside a result (relative change of $\bs$ w.r.t. 0-shot) indicates statistical significance compared to the relevant-only (oracle) method as per the paired $t$-test. Similarly, the significance of an IR model's ordered context (e.g., BM25) relative to its reversal (e.g., BM25-r) is indicated by `$\ddagger$'.
}

\small
\begin{adjustbox}{width=.95\textwidth}
\begin{tabular}{@{}l@{~~}l@{~~}l@{~~} cccc cccc@{}}
\toprule
& & & \multicolumn{8}{c}{Utility of RAG Context} \\
\cmidrule{4-11}
Query & 0-shot & IR & \multicolumn{4}{c}{$r_{\mathrm{DL}} \in \{2, 3\}$} & \multicolumn{4}{c}{$r_{\mathrm{DL}}=3,\, r_{\mathrm{dev\_small}}=1$} \\
\cmidrule(l){4-7} \cmidrule(l){8-11}
Set & $\bs$ & Model & $k=2$ & $k=5$ & $k=10$ & $k=15$ & $k=2$ & $k=5$ & $k=10$ & $k=15$\\
\midrule 

\multirow{6}{*}{DL'19} & \multirow{6}{*}{.6427} 
 & BM25 & .0582$\dagger$ & .1004 & .1065 & .0997 &  .0314$\dagger$ & .0798 & .0987 & .0901$\dagger$
\\
 & & BM25-r & .0736 & .1026 & .0902$\dagger$ & .0864$\dagger$ & .0123$\dagger\ddagger$ & .0844 & .0713$\dagger$ & .0698$\dagger$
 \\
 \cmidrule(r){3-11}
 & & MT5 & \textbf{.1161} & .0988 & .1201 & .1201 & .0809 & .0878 & .1010$\dagger$ & .0902$\dagger$ \\
 & & MT5-r & .1023 & .1149 & .0928 & .0996 & .0819 & .0966 & .0988$\dagger$ & .0886$\dagger$ \\
 \cmidrule(r){3-11}
 
 & & Rel & .1034 & \textbf{.1245} & \textbf{.1424} & \textbf{.1520} & \textbf{.1236} & \textbf{.1386} & \textbf{.1613} & \textbf{.1642}
\\
 & & NRel & -.0359$\dagger$ & -.0059$\dagger$ & -.0028$\dagger$ & .0013$\dagger$ & -.0082$\dagger$ & -.0256$\dagger$ & .0025$\dagger$ & -.0256$\dagger$\\
 
\midrule

\multirow{6}{*}{DL'20} & \multirow{6}{*}{.6159} 
 & BM25 & .0963 & .1204 & .1235 & .1333 & .1100$\dagger$	& .1298 & .1342 & .1424
  \\
 & & BM25-r & .0877 & .1139 & .0948$\dagger\ddagger$ & .0922$\dagger\ddagger$ & .0997$\dagger$ & .1047$\dagger$ & .0849$\dagger\ddagger$ & .0867$\dagger\ddagger$ \\
\cmidrule(r){3-11}
 & & MT5 & .1366 & \textbf{.1550} & \textbf{.1443} & .1323 & .1385 & .1537 & \textbf{.1565} & .1320$\dagger$ \\
 & & MT5-r & \textbf{.1526} & .1118$\dagger\ddagger$ & .1056 & .1066$\dagger$ & .1454 & .0953$\dagger\ddagger$ & .0857$\dagger\ddagger$ & .0866$\dagger\ddagger$\\
\cmidrule(r){3-11}
 & & Rel & .1401 & .1503 & .1359 & \textbf{.1602} & \textbf{.1675} & \textbf{.1627} & .1490 & \textbf{.1707}\\
 & & NRel & -.0224$\dagger$ & -.0147$\dagger$ & .0138$\dagger$ & -.0033$\dagger$ & -.0020$\dagger$ & -.0256$\dagger$ & .0043$\dagger$ & -.0031$\dagger$
\\

\midrule

 & \multirow{4}{*}{.5574} 
& BM25 & & & & & .0414 & .0617 & .0700 & .0721  \\
dev & & BM25-r & & & & & .0394 & .0567$\ddagger$ & .0623$\ddagger$ & .0580$\ddagger$ \\
\cmidrule(r){3-3} \cmidrule(r){8-11}
small & & MT5 & & & & & \textbf{.0915} &\textbf{.0944} & \textbf{.0965} & \textbf{.0916} \\
& & MT5-r & & & & & .0852$\ddagger$ & .0798$\ddagger$ & .0677$\ddagger$ & .0614$\ddagger$ \\
\bottomrule
\end{tabular}
\end{adjustbox}
\label{table:statistics}\vspace{-1em}
\end{table}

\section{Results}
In this section, we present our findings corresponding to the research questions as defined in Section \ref{sec:experiment}.

\para{RQ-1: Relation between retrieval effectiveness and downstream performance}


\begin{figure}[t]
\centering
\includegraphics[width=0.88\textwidth]{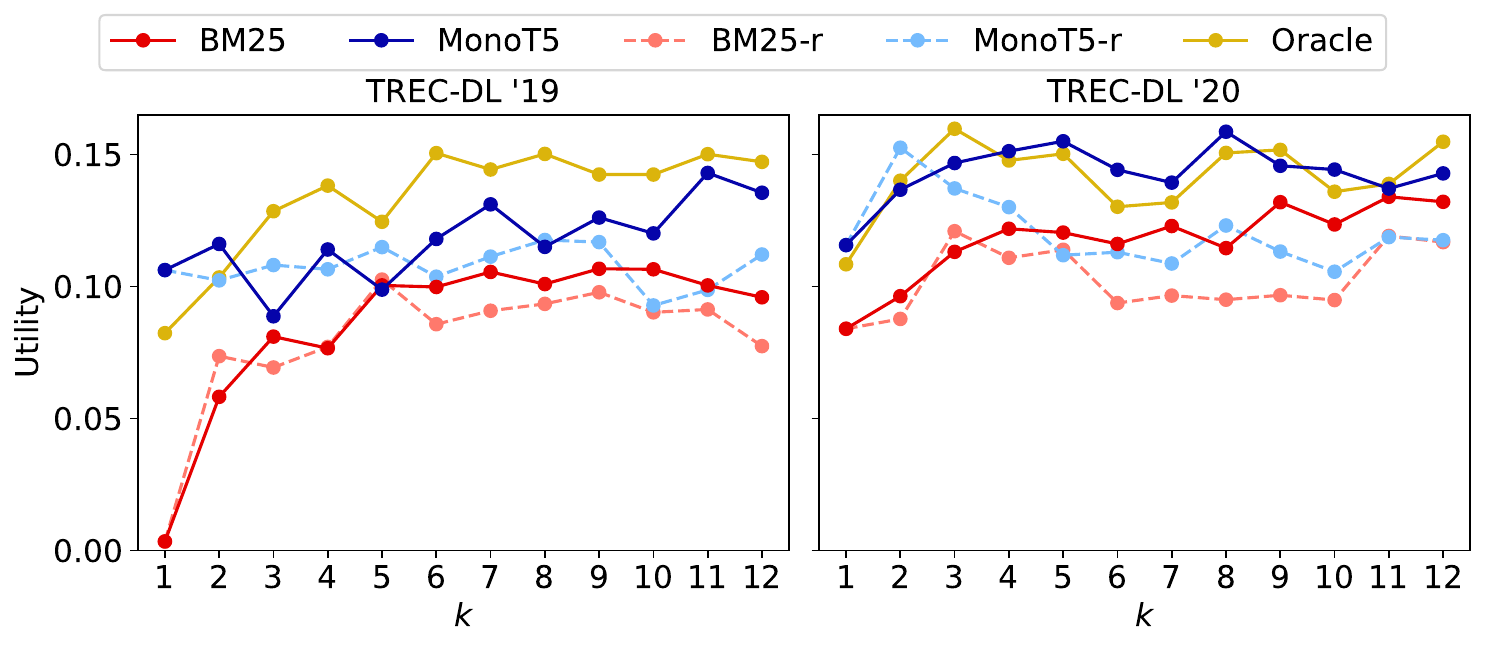}\vspace{-1em}
\caption{Average utility of the different RAG contexts with different sizes ($k$) for TREC DL-19 (left) and DL-20 (right). The relevant-only oracle method almost always yields the best utilities (only outperformed by MonoT5 for small context sizes). This indicates that it is not always the case that topical relevance of the context leads to better downstream performance. The utility values of the NRel-Only method (RAG context with non-relevant documents) are not shown in the plots as these values are significantly lower than all the other reported values.
}\label{fig:average_utility}\vspace{-1em}
\end{figure}

Table \ref{table:statistics} shows the average utility achieved by different RAG context sizes with two IR models (along with their reversal variants). We enlist the following important observations from Table \ref{table:statistics}.

\textit{First}, we observe that the context comprised of relevant documents only (Rel-Only) 
almost always achieves the highest utility, outperformed only by MonoT5 for small context sizes. These observations are consistent for two different ways of constructing the ground-truth to measure downstream effectiveness, i.e., both for $r_{\text{DL}}\geq 2$ and the more restrictive $r_{\text{DL}}=3$. Similar trends are also observed for dev\_small - a much larger set of queries with shallow relevance ground truth.

\textit{Second}, we observe that the more effective ranker, MonoT5, which is reported in the literature to yield much higher values of nDCG@$k$ than BM25 for the TREC DL benchmarks \cite{monot5}, also yields higher downstream utility than the weaker ranker (BM25). This shows the importance of employing a more effective ranking model as the `retriever' component of a RAG system. Nevertheless, despite being a relatively weak ranker, using BM25 as a traditional unsupervised model is still appealing because it leads to positive gains over 0-shot results, especially with larger context sizes.

\textit{Thirdly}, the results obtained with an effective IR model, such as MonoT5, are statistically indistinguishable from those achieved with a perfect ranker (paired $t$-test between them). This means that improving the topical relevance of retrieved contexts beyond a saturation point will likely have marginal incremental effects on RAG downstream effectiveness.

\textit{Finally}, the results with the reverse ordering of the top-$k$ documents indicate the importance of including relevant documents towards the top ranks. When context sizes are small (e.g., $k \leq 5$), 
\fzA{following or reversing the order induced by the retrieval scores does not significantly affect the utility, as shown by the fewer significant differences between them for smaller $k$s (viz. the number of $\ddagger$s present in Table \ref{table:statistics}).}
The most likely reason being that for an effective ranker the very top-retrieved documents are likely to be relevant. 
However, as expected, with an increase in context size, there is a higher likelihood of including non-relevant documents in the context, which means that ranking documents by their retrieval scores potentially plays a key role in determining the downstream task effectiveness.

\begin{figure}[t]
\centering
\includegraphics[width=0.8\columnwidth]{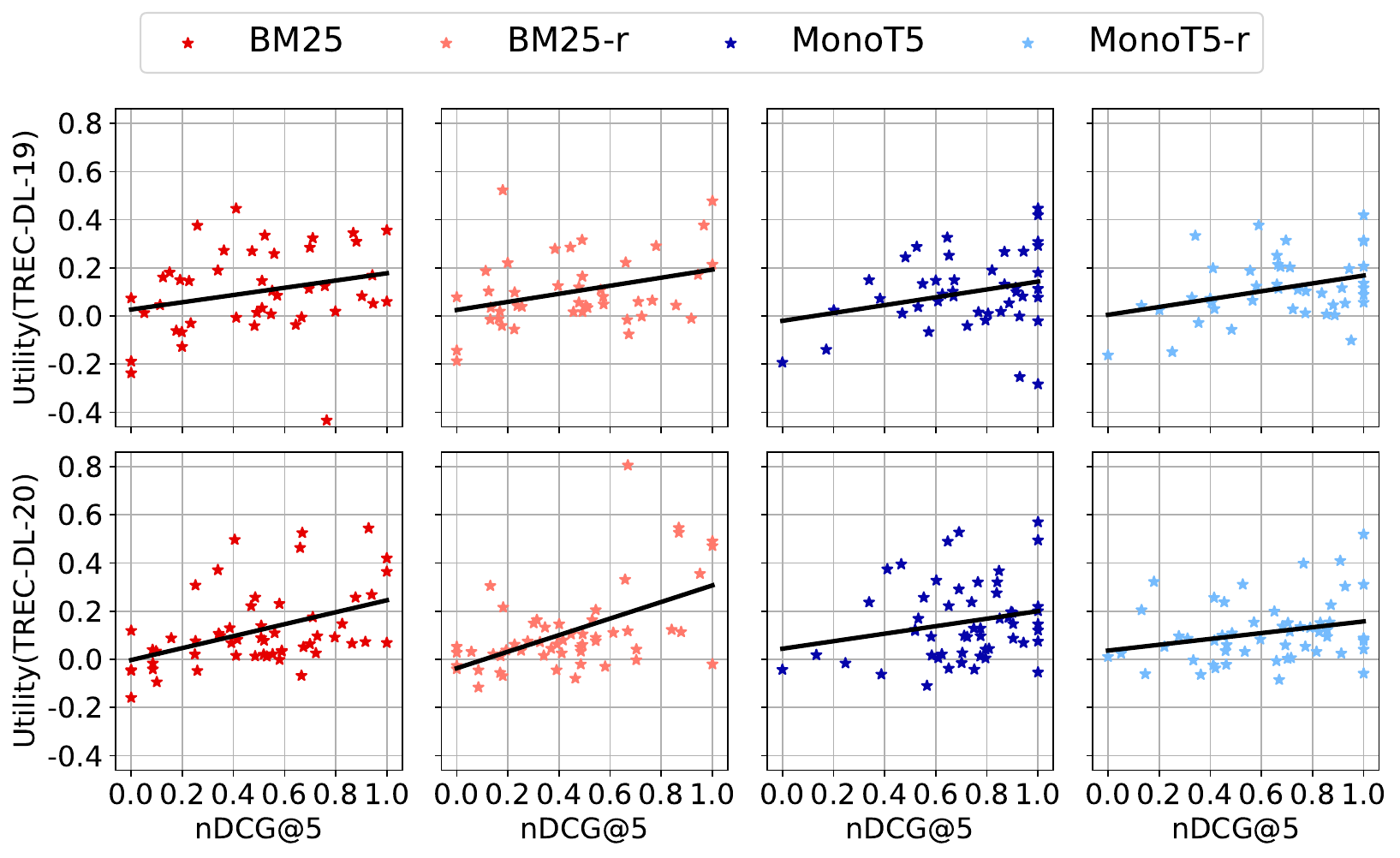}
\caption{Examples of the relationship between nDCG@$k$ and utility when $k=5$. The upper plots are for TREC DL-19, while the lower ones are for DL-20. There is generally a positive correlation between the effectiveness and utility, with different numbers of outliers in each subplot. Linear fits are shown as the black lines over the scatter plots.} \label{fig:ndcg_correlation_examples}\vspace{-1em}
\end{figure}

Figure \ref{fig:average_utility} presents a wider range of the spectrum of context size variations, which mostly corroborates the observations in Table \ref{table:statistics}. Comparing the two plots, it can be observed that the utility, generally speaking, also depends on the queries themselves. \cm{We observe} that the gains achieved by BM25-retrieved RAG contexts with smaller sizes are almost negligible for DL'19, which also suggests that the choice of the optimal IR model for a given query can potentially depend on latent factors such as query specificity~\cite{iclPerspective}.

\fzA{To conclude RQ-1, the contexts retrieved by more effective models yield higher utility. Reversing the order given by effective retrievers decreases utility.}

\begin{figure}[t]
\centering
\includegraphics[width=0.65\textwidth]{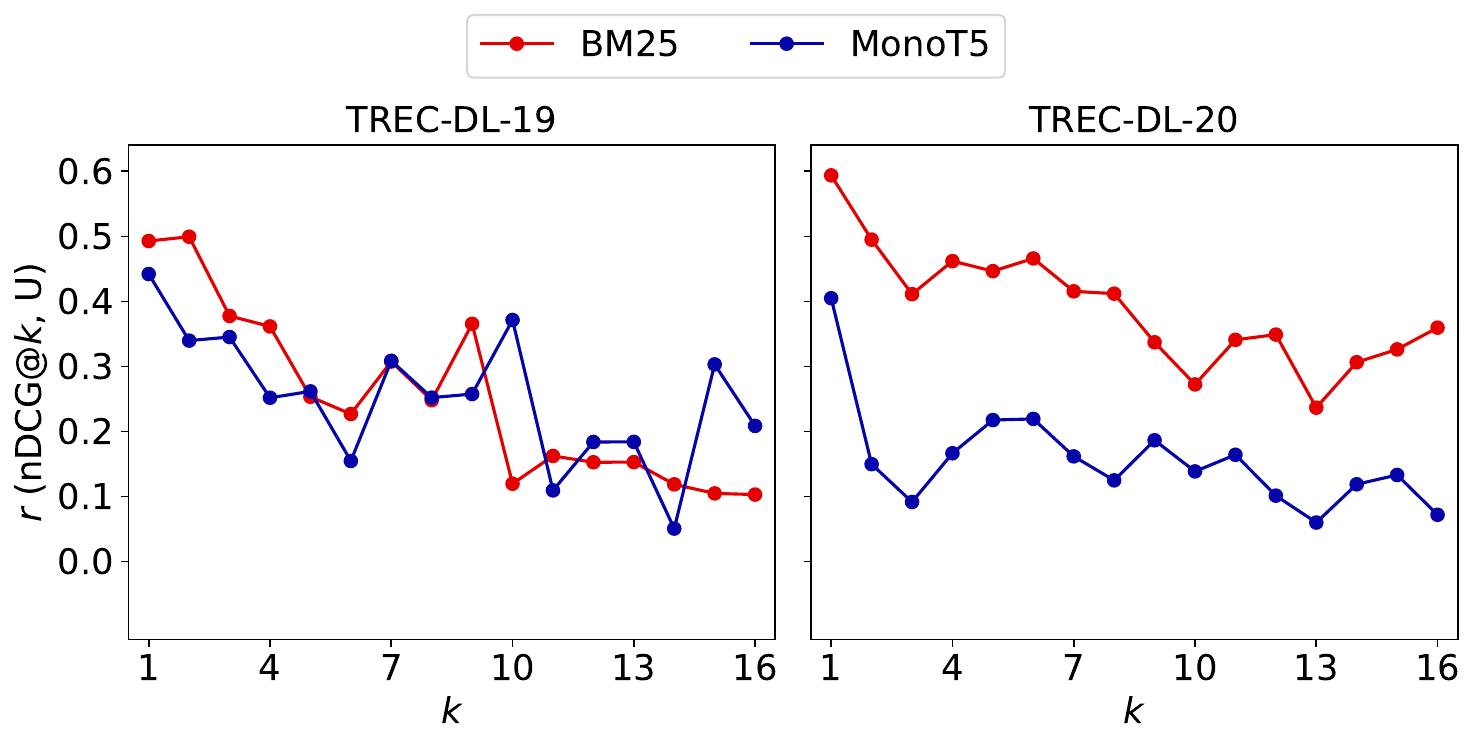}
\caption{The correlation between nDCG@$k$ and context utility in terms of Pearson's $r$ on TREC DL'19 and DL'20.
\label{fig:metrics_correlation_with_utility}
} 
\end{figure}

\para{RQ-2: Correlation values between contextual relevance and utility}


Figure \ref{fig:ndcg_correlation_examples} reports the correlation values as a scatter plot (Equation \eqref{eq:correlation} with Pearson's $r$) for a fixed context size of $k=5$. We report values for this particular value of $k$ as it belongs to neither of the two extremes - of too small or too large a context. We report the correlations for the two different retrieval approaches - BM25 and MonoT5 and their reversed permutations (for analysing the adversarial effects).

\looseness -1 From Figure \ref{fig:ndcg_correlation_examples}, we observe that there is a small positive correlation between relevance and utility: Pearson's $r$ can be visualised as the slope of the linear fits in the plots. The scatter plots show that a number of outliers exist for both IR models, i.e., where high relevance leads to low downstream performance. However, we find no evidence of low relevance leading to large gains in RAG performance.

Different from Figure \ref{fig:ndcg_correlation_examples}, which shows the per-query relevance-utility values for a fixed context size of $k=5$, Figure \ref{fig:metrics_correlation_with_utility} plots the relevance-utility correlation values over a range of context sizes. An interesting finding from Figure \ref{fig:metrics_correlation_with_utility} is that although the average utility keeps increasing with larger context sizes, the role of the relevance in ensuring a quality output keeps decreasing, reflected by the decrease in the correlation values. With higher context sizes, the likely presence of non-relevant documents potentially leads to this decrease in correlation between relevance and utility.

\fzA{To conclude RQ-2 - the correlation between the relevance of context and the corresponding utility is small but positive. It decreases as the number of documents in the context increases.}

\section{Concluding Remarks}

In this paper, we investigate the correlation between the topical relevance of a RAG context and its utility, thus capturing how effectively can the \textit{generator component} of a  RAG workflow translate the relevance, as obtained by its \textit{retriever component}, into a relevant answer at its output. The observed small positive correlations, which decrease as the context gets larger, indicate that relevance is not often and adequately transmitted from the retriever to the generated answers. This finding leaves the scope of devising novel retrieval models in the future, which will be aware of the downstream task objective in addition to topical relevance - an idea described in \cite{iclPerspective}.

\pageenlarge{1}An important difference of our evaluation compared to those of \cite{powerOfNoise,lostInTheMiddle,zamani-rag} is in the nature of the ground truth. While the notion of relevance in our setup is more robust (manually judged), the downside is that there is no manually formulated ground-truth answer as in the QA datasets \cite{kwiatkowski-etal-2019-natural,talmor-etal-2019-commonsenseqa}. With approximated ground truths for the generator, our observations on an IR test collection agree with the ones reported on KILT benchmarks \cite{powerOfNoise,lostInTheMiddle,zamani-rag} in the sense that more relevant RAG context
leads to better downstream performance. In contrast to the findings of \cite{powerOfNoise}, \cm{who found} that the downstream performance improves either when relevant documents are towards the top or the bottom of a ranked list, 
we observe that a reverse permutation of the context leads to a substantially detrimental effect, especially with large context sizes (see Table~\ref{table:statistics}).
Moreover, as the relevance of retrieved contexts in our setup is not simply `answer containment', it is difficult to construct a context that is `middle-heavy' in relevance, analogous to the ``Lost in the Middle'' paper \cite{lostInTheMiddle}. Our study, therefore, cannot comment on the effect of including relevant documents in the middle of a RAG context.

To address the limitation that our work considers approximated ground-truth of answers (as prescribed in \cite{benchmarkOfAnswerEvaluationFromRetriever}), we plan to work with manually generated reference answers for TREC DL queries. We would also like to apply the findings of this work to a more practical RAG workflow setup without the knowledge of any relevance assessments, which means that it is not possible to (exactly) compute the relevance of a RAG context. Query performance prediction (QPP)~\cite{DBLP:conf/wsdm/DattaGGM22,uef_kurland_sigir10,NQC} methods may then be applied to \textit{estimate the relevance of a RAG context} towards achieving per-query control of the RAG workflow \cite{DBLP:conf/ecir/DattaGMG24}.

\begin{credits}
\subsubsection{\ackname}
We would like to express our gratitude to the anonymous reviewers for their valuable feedback, which were helpful for us to improve this paper.

\subsubsection{\discintname}
There is no competing interest.
\end{credits}
%
%
%
%
\bibliographystyle{splncs04}
\bibliography{references.bib}
\end{document}